\documentclass[aps,prl,twocolumn,showpacs,groupedaddress]{revtex4}
\usepackage{graphics}
\begin{document}

\title{Saturation in heteronuclear photoassociation of $\bf^6$Li$\bf^7$Li}

\author{U. Schl\"oder, C. Silber, T. Deuschle, and C. Zimmermann}

\affiliation{Physikalisches Institut, Eberhard Karls Universit\"at, Auf der Morgenstelle 14, 72076 T\"ubingen, Germany}

\date{\today}

\begin{abstract}
We report heteronuclear photoassociation spectroscopy in a mixture of
magneto-optically trapped $^6$Li and $^7$Li. The laser-induced decrease in the
$^{7}$Li steady-state particle number, only appearing in the presence of $^6$Li,
gives clear evidence of photoassociation to form $^6$Li$^7$Li.
Hyperfine resolved spectra of the vibrational level $v\!=\!83$ of the singlet
state A$^1{\rm\Sigma}^+_u$ have been taken up to intensities of 1000\,W/cm$^2$. The
absolute resonance frequencies and the rotational constant have been measured. Saturation
of the photoassociation rate has been observed for two hyperfine transitions,
which can be shown to
be due to saturation of the rate coefficient near the unitarity
limit. Saturation intensities on the order of 40\,W/cm$^2$ can be determined.
\end{abstract}


\maketitle

\section{}
Photoassociation of cold atoms has been established as a versatile tool for the
formation of cold molecules. Starting with atoms either from a
magneto-optical trap or a Bose-Einstein condensate this technique has been
shown to allow for efficient production of various cold alkali dimers~\cite{fio,nik,gab,wyn}.
In single-photon photoassociation a pair of free atoms is optically excited
into a rovibrational level of the
electronically excited molecule. By spontaneous decay, either a pair of free
atoms is formed, or, with small fractional probability, a translationally cold ground state molecule. In extension to
this intrinsically incoherent process, also coherent processes, such as two-photon
Raman-type transitions~\cite{abr1,wyn,var,mac}, have been investigated. In the degenerate regime, such methods should allow for
coherent coupling between free atom-pair states and specific molecular ground
states, similar to the recently observed atom-molecule coupling due to a
Feshbach resonance~\cite{don}. It is expected that quantum statistics will affect the
initial atomic states as well as the molecular product, leading to what may
be phrased as `superchemistry'~\cite{hei}. In this context, mixtures between
bosonic and fermionic gases are particularly interesting, as they offer not only the
possibility of molecular Bose gases, but also of molecular Fermi gases.

A promising candidate for such experiments is lithium. It provides a stable
fermionic and bosonic isotope, it can be routinely cooled to quantum
degeneracy, and mixtures between Bose- and Fermi-gases have already been observed
and studied~\cite{tru,schr}. There is also extensive experience with photoassociation
of homonuclear lithium dimers in the nondegenerate and in the degenerate
regime~\cite{abr,ger}. However, the formation of fermionic molecules requires two
atomic species with different quantum statistics. In general, such
heteronuclear photoassociation bears intrinsic difficulties~\cite{sha} and,
although being very desirable as a universal tool~\cite{wei}, it has not yet been implemented successfully in binary
mixtures. For
studying heteronuclear photoassociation, lithium is again exceptionally
suitable. The isotopic shift of only 10\,GHz allows for resonant dipole-dipole interaction in the excited molecular state already at large internuclear separations, which is a crucial prerequisite
for efficient photoassociation~\cite{schl2}.

Even though the theory of photoassociation at high laser intensities is not yet fully
understood~\cite{boh,nap}, it is clear that for coherent coupling, optical excitation rates are required that are fast
as compared to the other relevant time scales of the system (i.e. collision
time and decoherence time due to spontaneous decay of the involved excited
states). One indication for entering the regime of sufficiently strong
excitation rates is the occurrence of saturation in the single-photon
photoassociation signal. Saturation studies have been reported for
several one-species experiments~\cite{dra,mck}. However, a clear saturation of the
photoassociation rate has not been observed yet.

In this paper we report on heteronuclear photoassociation in
a mixture of magneto-optically trapped $^6$Li and $^7$Li. We investigate photoassociation of
$^{6}$Li$^{7}$Li by monitoring the $^6$Li-induced decrease in the $
^{7}$Li steady-state particle number as a function of
the photoassociation laser frequency and intensity. The number of $^6$Li atoms
exceeds that of $^7$Li by a factor of 10 such that the photoassociation losses
are significant only for $^7$Li while the number of $^6$Li atoms remains almost
unaffected. This setup is possible only
for heteronuclear photoassociation and substantially facilitates the
quantitative analysis. Strong saturation of the photoassociation
rate is observed for two of four studied hyperfine lines of the chosen
singlet transition $v\!=\!83$.
We derive values for the saturation intensities on
the order of 40\,W/cm$^{2}$ which are sufficiently low to encourage
future experiments aiming at coherent control in the degenerate regime. 

Our experimental setup has been described in detail in~\cite{schl2}. The combined
magneto-optical trap (MOT) is operated
with two independent diode-laser systems and loaded from a single Zeeman-slowed atomic beam. The particle numbers are monitored by two absorption
lasers, tuned into resonance of the respective cooling transition, by means of a lock-in technique. The photoassociation light is provided by
a tunable dye laser, which is focused to a 1/e$^2$ diameter of 300\,$\mu$m,
resulting in peak intensities
up to 1000\,W/cm$^2$. Its frequency is measured with a combination of a commercial wavemeter, a Fabry-Perot etalon and a doppler-free iodine fluorescence spectroscopy~\cite{sor}. If the laser light is resonant with a molecular transition, losses in the
two-species MOT are induced due to spontaneous decay of the excited molecule into
a pair of free atoms or a ground state molecule. To simplify the quantitative
analysis, we operate the
MOTs such that the particle number of the $^6$Li MOT is large compared to that
of the $^7$Li MOT. Therefore, losses in the $^6$Li particle number due to
heteronuclear photoassociation are small and the $^6$Li MOT acts as a temporally constant
background gas for the $^7$Li MOT. Moreover, quadratic losses in the $^7$Li
particle number are
negligible. In that special case the
complicated solution of the coupled rate equations for a two-species system is reduced to the
following simple form for the $^7$Li steady-state particle number $N{_7}$:
\begin{equation}
N{_7}=\frac{L{_7}}{\alpha_{7}+\beta_{\rm PA}}.
\label{particlenumber}
\end{equation}
Here, $L{_7}$ is the $^7$Li MOT loading rate, and $\alpha{_7}$ is the loss rate
mainly due to collisions with hot atoms from the beam source. The
photoassociation rate $\beta_{\rm PA}=Kn_{6}wr$, with the photoassociation rate
coefficient $K$, the $^6$Li peak density $n_{6}$, the factor $w$ considering the
relative atomic hyperfine ground-state populations, and the factor $r$ accounting for
the spatial overlap between the atomic clouds and the laser
beam.  Spectra are taken by monitoring the $^7$Li steady-state atom
number as a function of the photoassociation laser frequency. With
1.25\,MHz/s, the scan rate was chosen slow as compared to the loading time of the
MOTs. The positions of the heteronuclear resonances can be predicted
theoretically within 1\,GHz, by applying the method of mass-reduced quantum
numbers~\cite{stw} to the well-known data for homonuclear
lithium~\cite{abr}. To identify
heteronuclear resonances we have recorded the $^7$Li particle number with and
without the other isotope present. The additional resonances give clear evidence of
$^6$Li$^7$Li photoassociation.

\begin{figure}
\hspace{-1.5cm}
\resizebox{0.5\textwidth}{!}{\includegraphics{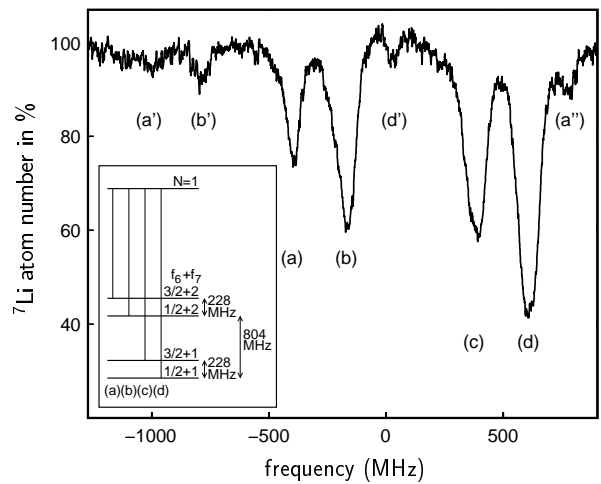}}
\caption{\label{hyperfine}Heteronuclear photoassociation spectrum of the
singlet $v\!=\!83$ level of $^6$Li$^7$Li. The frequency scale is relative to the
hyperfine center of gravity, which amounts to 14897.4015(7)\,cm$^{-1}$. The
inset shows the assignments of the different
hyperfine transitions.}
\end{figure}

We have concentrated on the transition into the vibrational level $v\!=\!83$
of the singlet series A$^1{\rm\Sigma}^+_u$ near a detuning of 177\,GHz to the
red of the
$D$1 line of $^6$Li. Its outer turning point is placed at an internuclear separation of about 74\,a$_0$. At this
detuning, the singlet series of $^6$Li$^7$Li is sufficiently separated
from the heteronuclear triplet resonances as well as from the homonuclear $^6$Li$^6$Li and
$^7$Li$^7$Li transitions, while at the same time the photoassociation rates are
still large. A high-resolution
photoassociation spectrum for an intensity of 420\,W/cm$^2$ is shown in
Fig.~\ref{hyperfine}. The four strongest lines ((a)-(d)) can be assigned to
transitions from the quartet of hyperfine entrance channels into the rotational
level $N\!=\!1$ of the excited state. This hyperfine structure
reflects the different combinations of atomic hyperfine ground states (hyperfine splitting:
228\,MHz for $^6$Li
and 804\,MHz for $^7$Li) and is a further clear indication of the
heteronuclear structure of the molecule. The absolute transition frequency,
adjusted to the hyperfine center of gravity, amounts to 14897.4015(7)\,
cm$^{-1}$. Several lines to the excited-state levels $N\!=\!0$
(resonances (a')-(d')) and $N\!=\!2$ (resonance (a'')) are also visible. From more sensitive spectra taken with a
less focused laser beam (1.0\,mm diameter), the rotational constant $B/h$ can be
determined as 301(2) MHz. The transitions into the $N\!=\!1$ level can be assigned to
$s$-wave collisions and the transitions into the levels $N\!=\!0,2$ to
$p$-wave collisions. The contributing partial waves are determined in their
parity by the selection rule on the change of angular momentum and in their number by the corresponding centrifugal barriers. For $d$-wave collisions the
barrier amounts to 38\,mK at an internuclear separation of
54\,a$_0$, which strongly suppresses contributions
from $d$-waves and above, even as shape resonances.

In the following we discuss the intensity dependence of the
photoassociation rate. We
took several series of spectra up to intensities of 1000\,W/cm$^2$ and find,
that the signal saturates. For all hyperfine transitions,
we observe a remaining fraction of $^7$Li atoms, that cannot be depleted, even in the
limit of high laser intensities. Thus, particle losses due to photoassociation
must be limited at high intensities. This limitation is to be explained
by the intensity dependence of the photoassociation rate. Internal MOT
processes, such as optical pumping, thermalization or diffusion, can be
excluded as the origin of the observed behaviour by a simple time scale
argument. In order to lead to a limited depletion, the losses due to photoassociation must occur on the same time scale as the loading time of the MOT, which is about
10\,s. Internal MOT processes happen on a much faster time scale and therefore
cannot limit the losses. 

For a quantitative analysis, one has to take into account that the
experimentally observed steady-state particle
number $N_{7}$ depends on the photoassociation rate in a nonlinear way (see Eq.~\ref{particlenumber}). We  
assume a Lorentzian frequency dependence of the
photoassociation rate $\beta_{\rm PA}=a/((f-f_{res})^{2}+\gamma^{2}/4)$, with the central frequency $f_{res}$, the linewidth $\gamma$ and a
proportionality factor $a$. Inserting this expression in
equation~\ref{particlenumber} yields:
\begin{equation}
N{_7}=\frac{N_{7,0}(1-cf)}{1+a/(\alpha_{7}((f-f_{res})^{2}+\gamma^{2}/4))}.
\label{fitfunction}
\end{equation}
We account for slow drifts in the loading rate during the frequency scan and
for partial overlap of the different lines by assuming a linear frequency dependence $\propto (1-cf)$ of the steady-state particle number without
photoassociation laser present $N_{7,0}$. For the atomic-beam limited loss rate
$\alpha_{7}$ we use the measured value of 0.11\,s$^{-1}$.
\begin{figure}
\hspace{-1.5cm}
\resizebox{0.5\textwidth}{!}{\includegraphics{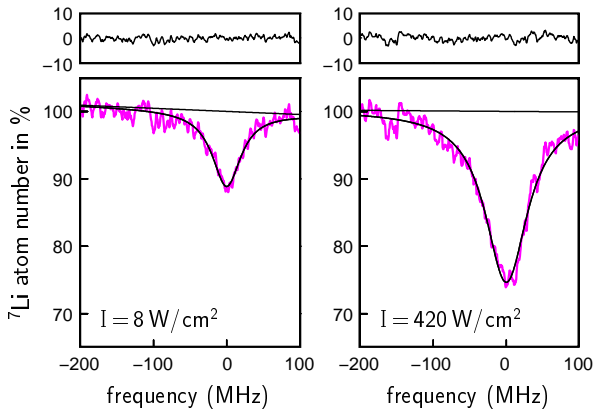}}
\caption{\label{peakfit}Spectra for the $f{_6}$=3/2\,+\,$f{_7}$=2 transition
(resonance (a)) at two different intensities. The solid line (lower panel) is a
fit according to Eq.~\ref{fitfunction}, the corresponding residuals are plotted
in the upper panel. The frequency scale is relative to the central frequency.}
\end{figure}
Up to intensities of 450\,W/cm$^2$ for
transitions (a) and (b) and of 600\,W/cm$^2$ for transitions (c) and
(d) the different hyperfine lines do not show any significant
asymmetries and partial overlap of the different lines remains small, so that
the resulting quasi-Lorentzian function represents the data sufficiently well. This is
exemplarily shown in Fig.~\ref{peakfit} for transition (a) at two intensities. Obviously, asymmetries due to thermal averaging
over the different collision energies~\cite{nap2} are negligible, since the natural linewidth of
the excited level (12\,MHz,~\cite{tho,cot}) plus the unresolved hyperfine-structure of the
excited state~\cite{tie} is larger than the thermal broadening of about 10\,MHz
for the MOT temperature of 0.5\,mK~\cite{schu}. 

\begin{figure}
\hspace{-1.8cm}
\resizebox{0.5\textwidth}{!}{\includegraphics{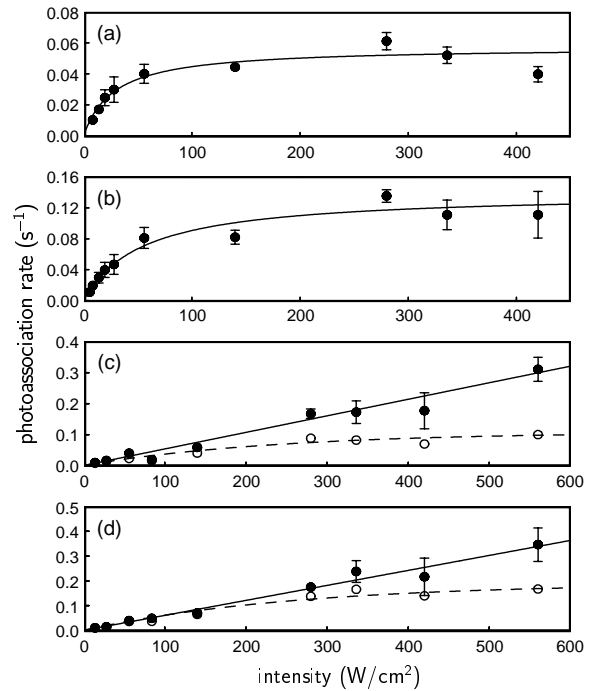}}
\caption{\label{saturation}Photoassociation rate as a function of the laser
peak intensity for the four different hyperfine transitions. The open circles
represent the uncorrected data.}
\end{figure}

For the determination of the photoassociation rate one has to realize that
the shape of the resonances can be influenced by various systematic effects. Due to the
focussing of the photoassociation laser and due to inhomogenities in the beam
profile, the light shift~\cite{boh} varies spatially, which leads to
inhomogeneous broadening. The different light shifts for the different
hyperfine levels of the excited state contribute similarily. These broadening
mechanisms lead to asymmetries at very high intensities and flatten the
approximately Lorentzian lines at moderate laser intensities. However, the area of the frequency-integrated
photoassociation rate is preserved, which has been checked numerically. Therefore, we correct the maximum photoassociation rate $4a/\gamma^2$,
which we get from the fit, by multiplying with the ratio of the broadened linewidth
$\gamma$ and the linewidth in the limit for zero intensity, $\gamma_{I=0}$,
which gives $\beta_{\rm PA,0}=4a/\gamma\gamma_{I=0}$~\cite{footnote}. This corrected
photoassociation rate $\beta_{\rm PA,0}$ is proportional to the area of the
rate and corresponds to the maximum rate in absence of any broadening mechanisms,
both broadening due to systematic effects and power broadening due to saturation. Extrapolation of the
broadened linewidths towards zero intensity leads to values for $\gamma_{I=0}$
of 51, 45, 33, 40\,MHz for resonances (a)-(d)  
respectively. They include the natural linewidth, the unresolved
structure of 
the different involved excited hyperfine states and thermal broadening. In
Fig.~\ref{saturation} the photoassociation rate for the four different
hyperfine transitions is presented as a function of the laser intensity. The data points are
mean values of up to 4 spectra. The error bars indicate the statistical error.
While the rates for the
transitions (c) and (d) do not saturate in the given intensity range (the
uncorrected rates $4a/\gamma^{2}$, which seem to saturate are also indicated), strong saturation is
observed for the transitions (a) and (b). As the $^6$Li density is nearly unaffected by
the photoassociation, this saturation must be due to saturation of the rate
coefficient $K$.

For the interpretation we use a
simplified model according to the close-coupled theory of Bohn and
Julienne~\cite{boh}. It yields an analytic expression for the rate coefficient
for a two-level system and a specific scattering energy. To compare with
the experimental corrected photoassociation rate, we have to adapt the
theoretical expression by multiplying  with the ratio of the power-broadened
linewidth and the natural linewidth. For $s$-wave
collisions this leads to the following intensity dependence for the corrected photoassociation
rate coefficient:
\begin{equation}
K_{0}=\frac{\pi v}{k^2}\frac{4}{1+I_{sat}/I}
\label{intensity}
\end{equation}
where $v$ is the velocity and $k$ is the wave number of the relative motion. $I_{sat}$ is the energy-dependent saturation intensity, for
which the natural linewidth equals the induced linewidth. This intensity dependence also applies for the
thermally-averaged corrected rate coefficient with
approximately the same saturation intensity, as checked by simulations. A full theoretical
analysis would include a population-weighted summation over the different
transitions from the magnetic ground-state sublevels to the various
excited hyperfine states and has not been carried out yet.
We thus assume the
validity of the expression (Eq.~\ref{intensity}) also for the full problem, and
use it for a fit to the data of the saturating transitions. For the saturation
intensities we obtain values of 28\,W/cm$^2$ for transition (a) and of
54\,W/cm$^2$ for transition (b). This is about
10$^4$ times the atomic saturation intensity and in good agreement with an estimation for a two-level system, which gives 33\,W/cm$^2$
for an energy of $E/k_{\rm B}\!=\!0.5$\,mK~\cite{jul}.
  For the corrected photoassociation rates, we receive maximum values of 0.06\,s$^{-1}$ for resonance (a) and
0.14\,s$^{-1}$ for resonance (b). Taking into account an estimated $^6$Li peak intensity of
10$^{10}$\,cm$^{-3}$, the overlap factor $r\!=\!0.22$, the population factor
$w$ (0.42 and 0.21 for the transitions (a) and (b) respectively, statistical
distribution assumed), and the correction factor of 4 (see Eq.~\ref{intensity}), the saturated photoassociation rate coefficient $K$ can be
estimated to be (2$\cdot$10$^{-11}-8\cdot$10$^{-11}$)\,cm$^3$s$^{-1}$. This
value is close to the unitarity limit, where the scattering probability $\vert S\vert^2\rightarrow$ 1. In this limit the rate coefficient is simply
given by the product of the scattering cross-section $\pi/k^2$ and the velocity
$v$, resulting in 5$\cdot$10$^{-10}$\,cm$^3$s$^{-1}$ for the temperature considered. For a more sophisticated
analysis, knowledge of the different ground-state sublevel populations, of the
various Clebsch-Gordan coefficients and of laser-polarization effects would be required.
 
In summary, we have observed heteronuclear photoassociation in
$^6$Li$^7$Li. Studying the intensity dependence of the hyperfine resolved
singlet spectrum, we have observed saturation of the rate coefficients for two
of the four hyperfine lines and determined the corresponding saturation
intensities. The observation of the unitarity limit has been possible, as it is
orders of magnitude lower at MOT temperatures than at temperatures in
the degenerate regime~\cite{mck}. Reaching the unitarity limit is another proof
that already at MOT temperatures quantum mechanical properties become apparent.

\begin{acknowledgments}
We are grateful to P.S. Julienne and E. Tiemann for enlightening discussions and
theoretical support. This work has been partially funded by the Deutsche Forschungsgemeinschaft.

\end{acknowledgments}

%%%%%%%%%%%%%%%%%%%%%%%%%%%%%%%%%%%%%%%%%%%%%

%\begin{thebibliography}{}
%\bibliography{neubau}

\begin{thebibliography}{29}
\bibitem {footnote}
\eprint{For the analysis we follow [20]. In contrast to the simulation
mentioned in this reference, our
simulation shows, that the frequency-integrated photoassociation rate is always
slightly larger with the various systematic effects considered than
without.}
\end{thebibliography}

%%%%%%%%%%%%%%%%%%%%%%%%%%%%%%%%%

\end{document}